\documentclass[apj]{emulateapj}
\usepackage{amsmath}
\usepackage{amssymb}
\usepackage{epsfig}
\usepackage{apjfonts}

\newcommand{\cms}{\,{\rm cm$^{-2}$}\,}
\newcommand{\cmc}{\,{\rm cm$^{-3}$}\,}
\newcommand{\kms}{\,{\rm km\,s$^{-1}$}\,} 
\newcommand{\kmsmpc}{\,{\rm km\,s$^{-1}$\,Mpc$^{-1}$}\,}

\newcommand{\etal}{{ et~al.~}}

\newcommand{\ergs}{\,{\rm erg\,s$^{-1}$}\,}

\newcommand{\Ms}{M_\odot}
\newcommand{\Zs}{Z_\odot}

\newcommand{\cnts}{\,{\rm cnts\,s$^{-1}$}\,}
\shorttitle{Active Nucleus of IC~4970}

\begin{document}


\title{The Active Nucleus of IC~4970: A Nearby Example of 
 Merger-Induced Cold-Gas Accretion }

\author{M.E. Machacek, R.P. Kraft, M.L.N. Ashby, D.A. Evans, C. Jones, \&
 W.R. Forman} 
\affil{Harvard-Smithsonian Center for Astrophysics \\ 
       60 Garden Street, Cambridge, MA 02138 USA}
\email{mmachacek@cfa.harvard.edu} 

\begin{abstract}

We present results from {\it Chandra} X-ray and {\it Spitzer} 
mid-infrared observations of the interacting galaxy pair 
NGC~6872/IC~4970 in the Pavo galaxy group and show that the smaller 
companion galaxy IC~4970 hosts a highly obscured active galactic 
nucleus (AGN). The $0.5-10$\,keV X-ray luminosity of the nucleus is variable, 
increasing by a factor $2.9$ to $1.7 \times 10^{42}$\ergs 
(bright state) on $\sim 100$\,ks timescales. The X-ray spectrum of the 
bright state is heavily absorbed ($N_{\rm H} = 3 \times 10^{23}$\cms for 
power law models with $\Gamma = 1.5-2.0$) and shows a clear 
$6.4$\,keV Fe K$\alpha$ line with equivalent width of $144-195$\,eV.
Limits on the diffuse emission in IC~4970 from 
{\it Chandra} X-ray data suggest that the available power from Bondi
accretion of hot interstellar gas may be an order of magnitude too small
to power the AGN. {\it Spitzer} images show that $8\micron$ nonstellar 
emission is concentrated in the central $1$\,kpc of IC~4970, 
consistent with high obscuration in this region. The mid-infrared colors 
of the nucleus are consistent with those expected for a highly 
obscured AGN. Taken together these data suggest that the nucleus 
of IC~4970 is a Seyfert $2$, triggered and fueled by cold material 
supplied to the central supermassive black hole as a result of 
the off-axis collision of IC~4970 with the cold-gas rich spiral galaxy 
NGC~6872. 

\end{abstract}



\section{INTRODUCTION}
\label{sec:introduction}

Supermassive black holes are known to exist at the centers of 
most galaxies. The strong correlation of the central supermassive
black hole mass with the mass of the host galaxy's stellar bulge, 
as represented 
by the central stellar velocity dispersion (Ferrarese \& Merritt
2000;  Gebhardt \etal 2000) or  K-band  absolute magnitude 
(Marconi \& Hunt 2003), indicates that the evolution of the central 
black hole and that of its host galaxy are inextricably linked. 
Most recent work relating the activity of the central supermassive
black hole to galaxy 
evolution has focused on the interaction of 
radio-loud active galactic nuclei (AGN) with their environment, in
order to resolve the cooling flow problem in galaxy clusters and bring 
hierarchical models of galaxy formation into agreement with 
observations of galaxies at low redshift
(see, e.g., Dunn \& Fabian 2006; Best \etal 2006; Croton \etal 2006). 
Hardcastle \etal (2007) separated these AGN into two classes, 
low excitation and high excitation radio galaxies, based
on their optical emission line characteristics, 
and suggested that the observed differences between the two classes 
are determined by the fuel source and accretion mode of the 
central black hole. 
  Low excitation radio galaxies tend to be massive, 
dominant elliptical galaxies, residing in the 
deep dark matter potentials near the cores of large groups and rich 
clusters, and are surrounded by an ample supply of hot gas from the 
 intracluster medium falling  
into their large gas halos. They show no evidence for a standard,
geometrically thin, optically thick, luminous accretion
disk. Radiatively inefficient Bondi accretion of 
hot gas accreted from their surroundings onto the central supermassive
black hole is sufficient to supply the total 
energy, mechanical as well as radiative, observed in these systems. 
In contrast, high excitation radio galaxies (containing the most
powerful radio galaxies known) 
exhibit high-excitation, narrow-line
optical emission in their spectra, characteristic of emission from 
an accretion disk. Hardcastle \etal (2007) argue that AGN activity in 
these galaxies is powered by radiatively efficient accretion 
of cold gas acquired in a major merger with a massive cold-gas-rich
partner. The high radio power and jets observed in high excitation
radio galaxies often may result 
from the formation of a rapidly spinning nucleus
from the coalescence of the merging galaxies' two massive black holes 
(Wilson \& Colbert 1995; Capetti \& Balmaverde 2005). 

Since cold-gas accretion is associated with the galaxy's interaction
and merger history, and does not depend on the depth of the dark matter
potential or the presence of a hot intracluster medium,  
galaxy collisions of lower mass galaxies with dusty, gas-rich partners 
in galaxy groups also would be expected to trigger AGN activity and 
black hole growth. Gravitational and
hydrodynamical forces, induced by the galaxies' mutual interaction, funnel 
dust and cold gas from the gas-rich partner into 
the interacting, lower-mass companion galaxy's nuclear region, where it settles
into an optically thick accretion disk before subsequently  
accreting onto the central black hole. As in the
evolutionary model of Churazov \etal (2005), accretion onto the 
supermassive black hole is radiatively efficient and radio emission 
from the AGN may be weak or absent. Such galaxy collisions were likely  
frequent at high redshift ($z \sim 1.3$), when the fraction of blue, 
gas-rich starforming galaxies in moderately massive galaxy groups was
high and galaxies were rapidly evolving (Cooper \etal 2006; 
Gerke \etal 2007). Thus, while perhaps less dramatic than radio-loud
systems, cold-gas accretion in off-axis galaxy collisions and minor mergers
may play an equally important role in the evolution of lower
mass galaxies and moderately massive galaxy groups into what we see
today.

\begin{figure}[t]
\begin{center}
\includegraphics[height=2.0in,width=3in]{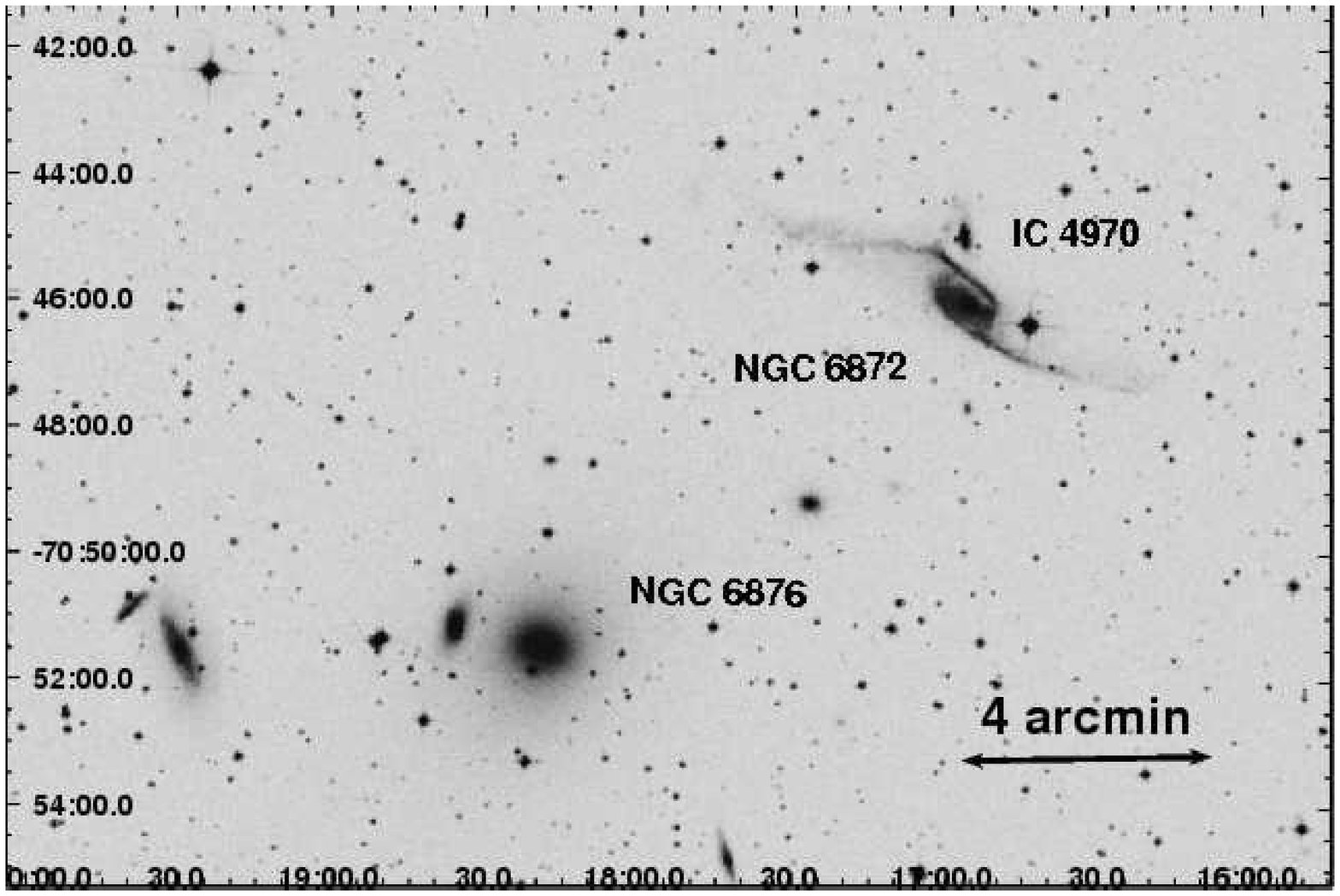}
\includegraphics[height=2.0in, width=3in]{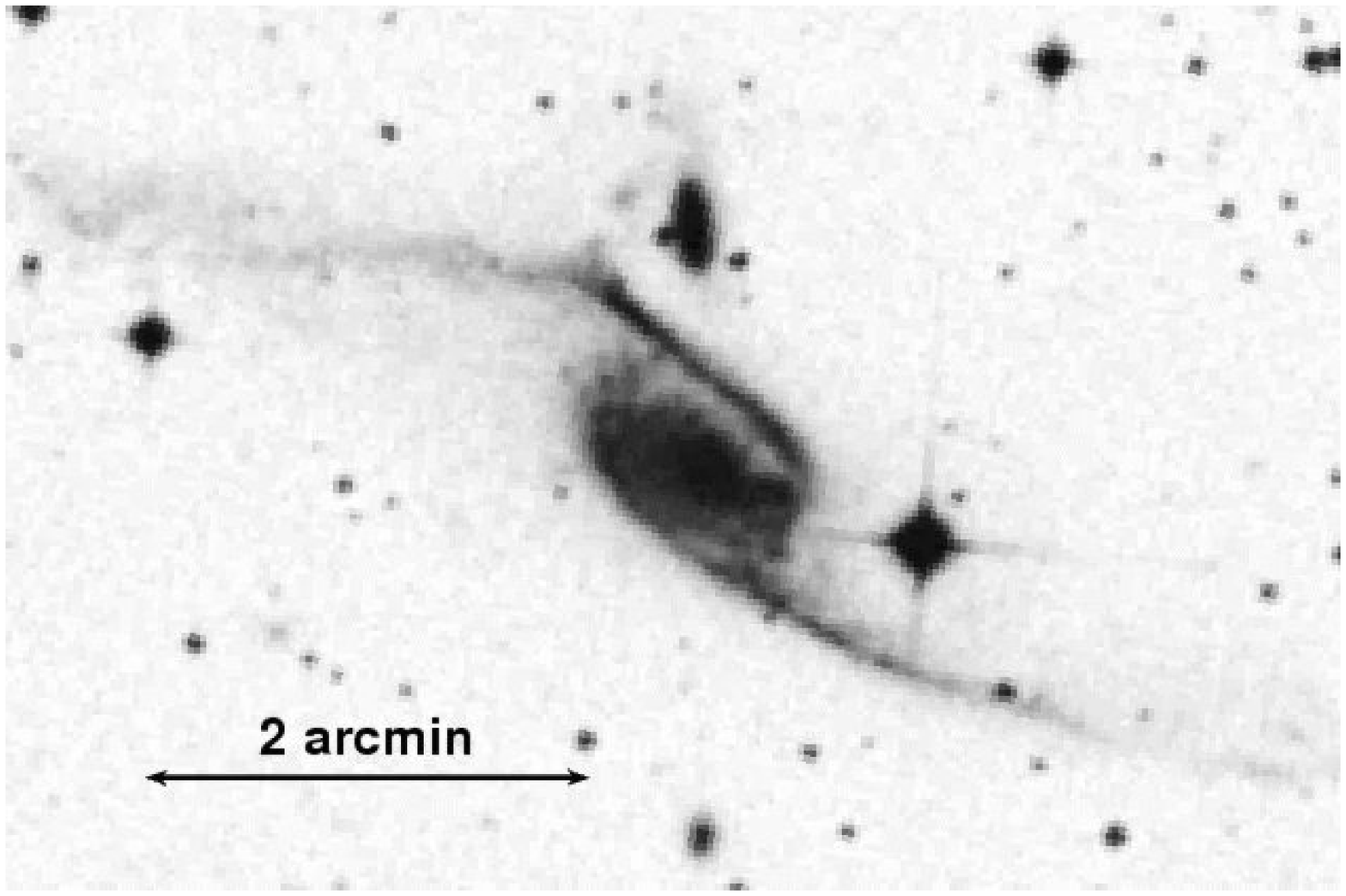}
\caption{({\it top}) Digitized Sky Survey  Bj band image, taken 
  with the UK $48$inch Schmidt Telescope, of the dominant elliptical galaxy
  NGC~6876 and nearby NGC~6877 at the center of the Pavo group  and 
the large spiral galaxy NGC~6872 with its tidally 
interacting companion galaxy IC~4970 located $8\farcm7$ to the northwest.
({\it bottom}) Zoom-in from the image in the top panel to highlight the 
 interaction of IC~4970 with the spiral galaxy NGC~6872. 
}
\label{fig:pavogroup}
\end{center}
\end{figure}

Observations of nearby interacting galaxies in similarly moderately 
massive groups offer a unique window into the dynamical processes 
that triggered AGN activity in lower mass galaxies  
and influenced the coevolution of the host galaxy with its central
supermassive black hole at this earlier epoch, when galaxies were 
rapidly transforming. 
The Pavo group is such a nearby ($z=0.01338$) group 
with a cool $0.5$\,keV intra-group medium, $\sim 13$ member galaxies and 
velocity dispersion $\sim 425$\kms (Machacek \etal 2005) similar to
the properties of $z \sim 1$ groups observed in the DEEP2 survey 
(Gerke \etal 2005).
In the top panel of Figure \ref{fig:pavogroup}, we present a Bj-band 
image of the dominant Pavo group galaxies, showing the elliptical 
galaxy NGC~6876 at the Pavo group 
center and the large, gas-rich
tidally-distorted SAB(rs)c spiral galaxy NGC~6872, located $8\farcm7$ to
the northwest. The  smaller lenticular galaxy IC~4970, that is only $1/5$ to
$1/10$ as massive as the spiral galaxy NGC~6872 (Mihos \etal 1993), 
is located $1\farcm12$ to the north of NGC~6872's center   
near a break (`knee') in NGC~6872's northern tidal arm. 
The small $74$\kms line-of-sight velocity difference\footnote{All 
line-of-sight velocities in this paper are taken from the CfA Redshift 
Survey by Martimbeau \& Huchra. Online versions of the catalog, supporting
software, and documentation on the CfA Redshift Survey are available
at http://cfa-www.harvard.edu/\~\,huchra/zcat.} 
between IC~4970 and NGC~6872 suggests that
the galaxy pair forms a spiral-dominated subgroup, while    
{\it XMM-Newton} observations of a hot, gas trail linking, in
projection, the central elliptical galaxy NGC~6876 and
NGC~6872/IC~4970 indicate that 
the NGC~6782/IC~4970 subgroup has recently passed supersonically
through the Pavo group core (Machacek \etal 2005).
Thus the Pavo group, like groups at 
high redshift, may be dynamically young and the interactions between
galaxies representative of those important to galaxy evolution at that
earlier epoch. 

IC~4970 and NGC~6872 have long been known to be a tidally interacting 
galaxy pair (Vorontsov-Velyaminov 1959). As shown in the bottom panel
of Figure \ref{fig:pavogroup}, tidal tails extend outward 
more than $30$\,kpc ($\sim 2\farcm$) to the east and west from the tip of 
NGC~6872's spiral arms. A stellar bridge connects the 
`knee' in NGC~6872's northern 
arm and tidal tail to the lenticular companion galaxy IC~4970, 
indicating that tidal interactions between IC~4970 and NGC~6872 are
ongoing. The spiral galaxy NGC~6872 is gas rich. From CO J=1-0 line
emission, Horellou \& Booth (1997) infer  
$9.6 \times 10^8\Ms$ of molecular hydrogen in NGC~6872's central region.
Although atomic hydrogen is not observed near the center 
of the spiral galaxy,  $1.41 \times 10^{10}\Ms$ of HI gas is 
distributed along the full extent of NGC~6872's spiral arms and 
associated tidal tails (Horellou \& Booth 1997; 
Horellou \& Koribalski 2007) and  $1.3 \times 10^9\Ms$
of HI gas is seen along the direction of IC~4970, that may be
associated with the companion galaxy (Horellou \& Koribalski 2007). 
H$\alpha$ emission traces ionized gas from recent star formation
along NGC~6872's outer spiral arms to the ends of the tidal tails and
from the `knee' north in the stellar bridge (Mihos \etal 1993). 
The distribution of H$\alpha$ 
corresponds closely to the distribution of young star clusters 
observed in this system (Bastian \etal 2005). However, 
Mihos \etal (2003) found  no H$\alpha$ emission from the central 
regions of NGC~6872 or associated with the companion galaxy IC~4970.

In this paper we focus on the effects of the IC~4970/NGC~6872 
collision on the lower mass participant IC~4970, based on 
{\it Chandra} X-ray
and {\it Spitzer} mid-infrared  observations that show IC~4970 
hosts a highly-obscured active nucleus. Results for the two 
massive galaxies, NGC~6872 and NGC~6876, and the Pavo group gas  
from these observations will be reported in 
subsequent papers (Machacek \etal 2007a,b in preparation).
Our discussion is organized as follows:  
In  \S\ref{sec:obs} we briefly review the {\it Chandra} and {\it
  Spitzer} observations and our data reduction and 
  processing procedures. 
In \S\ref{sec:analysis} we present our main analysis results.
 We discuss the X-ray time variability of the nuclear point source
 in \S\ref{sec:vary}, its X-ray spectral properties in 
 \S\ref{sec:xrayspec}, and mid-infrared flux densities and colors 
 in \S\ref{sec:iraccol}, and show
 that these properties are consistent with the classification of IC~4970's
 nucleus as a highly obscured Seyfert $2$. In \S\ref{sec:ULX} we
 briefly discuss the properties of a nearby ultra-luminous X-ray
 source  and in 
\S\ref{sec:diffuse} place limits on the amount of 
hot gas outside the central region of the galaxy.
 In \S\ref{sec:discuss} we determine the
mass of the central black hole in IC~4970 and discuss possible
accretion modes, concluding that nuclear activity in
IC~4970 is most likely triggered and fueled by cold gas driven into 
the nucleus during IC~4970's ongoing  off-axis collision with the dust- and
gas-rich spiral galaxy NGC~6872. We summarize our
results in \S\ref{sec:conclude}. 
Unless otherwise indicated, 
we quote $90\%$ confidence levels for the uncertainties in spectral 
parameters and $1\sigma$ uncertainties on counts and count rates. 
Coordinates are J2000.0. Adopting the flat $\Lambda$CDM cosmology 
from the three year WMAP results ($H_0 = 73$ \kmsmpc, $\Omega_m =
0.238$, Spergel \etal 2007), the luminosity distance to the Pavo Group 
($z=0.01338$) is $55.5$\,Mpc and $1\farcs$ corresponds to a distance scale
of $0.262$\,kpc.


\section{OBSERVATIONS AND DATA REDUCTION}
\label{sec:obs}

\subsection{{\it Chandra} Observations}
\label{sec:chandraobs}

Our X-ray data consist of two observations   
of the Pavo galaxy group, including IC~4970, taken with the 
ACIS-I detector (Garmire \etal 1992; Bautz \etal 1998) on board 
the {\it Chandra} X-ray Observatory 
on 2005 December 14-15 (OBSID 7248) and 2005 December 16-17 (OBSID
7059) with total exposures of $35.3$ and $40.7$\,ks, respectively. 
The nominal ACIS-I aimpoint was adjusted $-2\farcm0$ in the negative 
Y-axis direction away from the nucleus of the spiral galaxy NGC~6872 
to prevent chip gaps from falling on the tidal features and diffuse 
emission regions of interest. The data were analyzed using standard 
X-ray analysis packages CIAO3.3, FTOOLS, and XSPEC 11.3.0. 
The data were filtered to remove 
events with bad grades ($1$, $5$, $7$), and those that fell on hot
pixels. Events, flagged in VFAINT mode as having excessive flux in the 
border pixels surrounding event islands were also removed, resulting
in a factor $\gtrsim 2$ improvement in the rejection of particle 
backgrounds at energies below $\sim 1$\,keV. We then reprocessed 
the data and created response
files using the most recent gain tables and instrumental corrections, 
including corrections for the charge transfer inefficiency  of
the ACIS-I forward illuminated CCDs, the time-dependent build-up of 
contaminants on the telescope optical filter (Plucinsky \etal 2003),  and the 
secular drift of the average pulse height amplitude for photons of fixed
energy (tgain).\footnote{See http://cxc.harvard.edu/contrib/alexey/tgain/tgain.html}
Periods of anomalously high and low count rates were removed from each
observation using the script lc\_clean acting on events in the 
$0.3-12$\,keV energy band from the I0 CCD, where there were no bright 
sources. This resulted in useful exposures of $31,878$\,s for OBSID
7248 and $38,364$\,s for OBSID 7059. 

Background files for image
analysis were generated using the $1.5 \times 10^6$\,s source free 
data set acisi\_D\_01236\_bg\_evt\_010205.fits, appropriate for the 
date and instrument configuration of our 
observation.\footnote{see http://cxc.harvard.edu/contrib/maxim/acisbg} 
The background files were normalized by comparing rates in the 
$9.0-11.5$\,keV energy band, where particle background dominates. 
We found that the
$9.0-11.5$\,keV event rates in our observations were $18.2\%$ ($18.9\%$) higher
 in OBSID 7248 (7059), respectively, than that found in the source
 free datasets. We thus applied an additional scaling factor of $1.182$ 
($1.189$) to the source free data sets for OBSID 7248 (7059),
 respectively, to bring the $9.0 - 11.5$\,keV source free event rates 
into agreement with the observations. 

\subsection{{\it Spitzer} Observations}
\label{sec:iracobs}

We observed NGC~6872/IC~4970 in the mid-infrared ($3.6$, $4.5$, $5.8$, and 
$8.0\micron$ wavebands) on 2005 September 18 using the 
Infrared Array Camera 
(IRAC; Fazio \etal 2004; Hora \etal 2004) on board the Spitzer Space 
Telescope (Werner \etal 2004), 
as part of a $432$\,s  {\it Spitzer} IRAC observation (AORID 14699264) 
of the Pavo Galaxy Group.
Our IRAC observation of the Pavo group used two fixed
positons to ensure coverage of both the central group elliptical
galaxy NGC~6876 and the NGC~6872/IC~4970 interacting galaxy pair. We 
used  the $12$\,s HDR mode and  a $9$-position, large-scale random 
dithering pattern 
to minimize  latency artifacts, while providing sufficient redundancy 
for reliable removal of cosmic rays and scattered light
artifacts. This resulted in an effective integration time at the 
position of IC~4970 of $108$\,s 
in the $3.6$ and $5.8\micron$ wavebands and $216$\,s 
in the $4.5$ and $8.0\micron$ wavebands.  
We used Basic Calibrated Data  reprocessed by the Spitzer Science 
Center S14.0 pipeline 
on 31 May 2006 to include the super-boresight correction for 
improved astrometry. The Basic Calibrated Data frames in all 
four IRAC wavebands 
were cleaned using the script cosmetic.pl in the  
Spitzer Science Center data analysis software suite MOPEX (version
030106) to mitigate muxiplexer bleed, and to correct for column 
pulldown and dropouts. 
We then used the `post-Basic Calibrated Data' processing suite 
IRACproc 4.1.2 (Schuster \etal 2006)
with linear interpolation and $0\farcs86$ pixel size to reject outliers
and prepare mosaics, registered on the same sky grid, in each
waveband. A small, residual background gradient persisted across the
mosaics in each channel. We modeled this background gradient using 
the SExtractor 2.5.0 software package (Bertin \& Arnouts 1996). All of  
the resulting background maps were inspected and found free of anomalies
that would contaminate the photometry. We then subtracted the   
background map for each channel from the data to create the final
background-subtracted mosaics. 

\section{Data Analysis}
\label{sec:analysis}

\begin{figure}[t]
\begin{center}
\includegraphics[height=1.89in,width=3in]{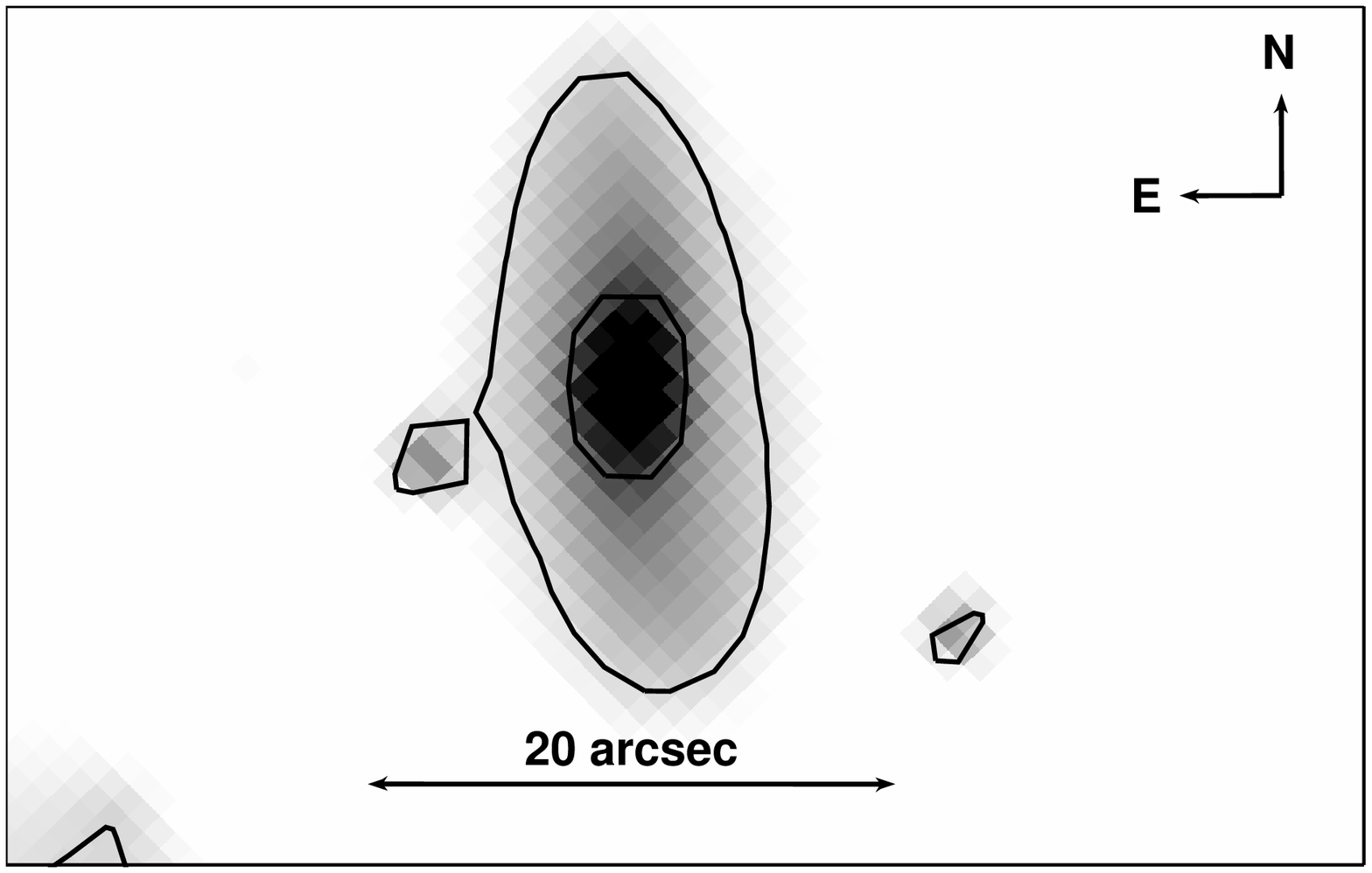}
\includegraphics[height=1.89in,width=3in]{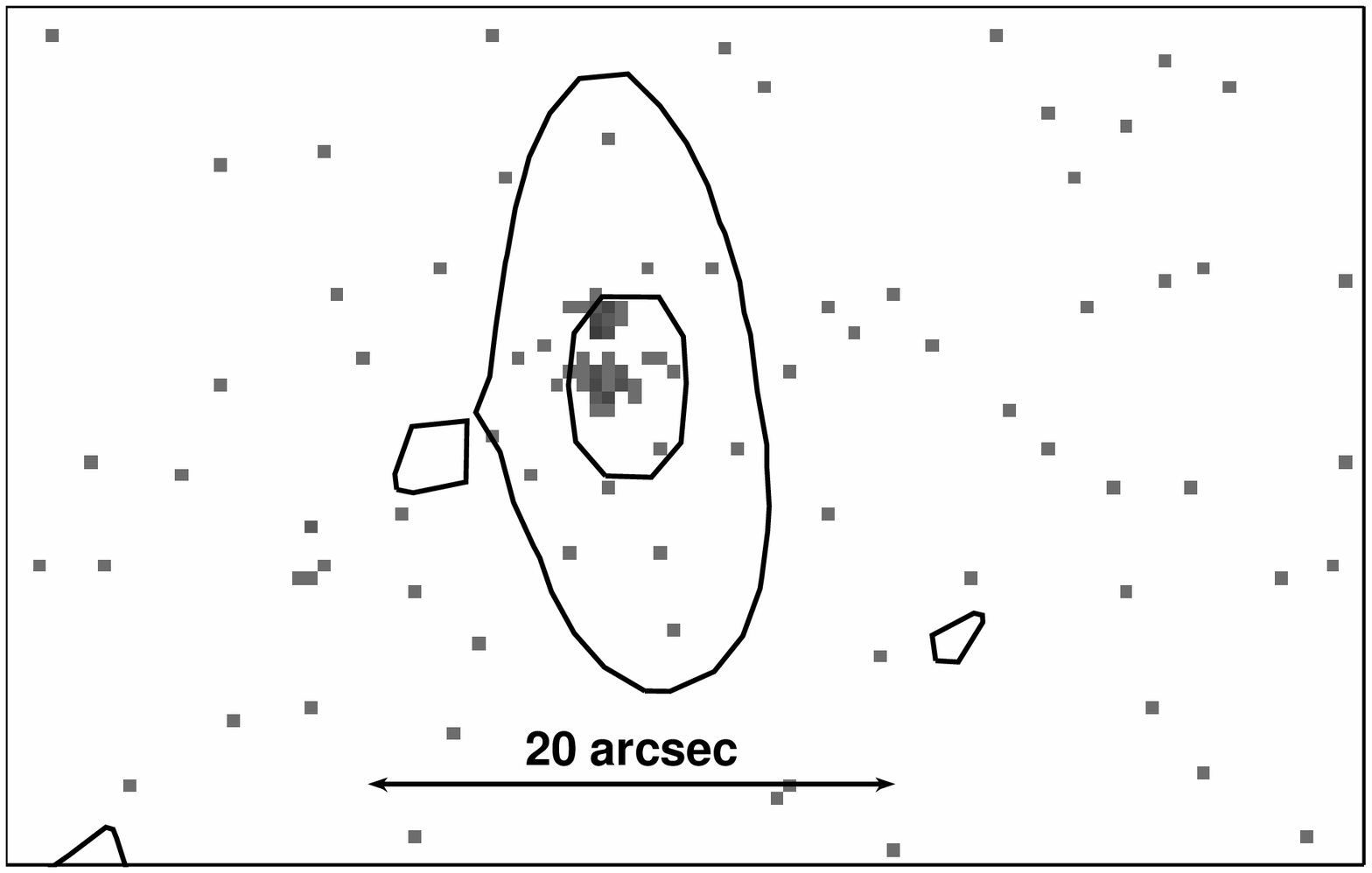}
\caption{ ({\it top}) {\it Spitzer} IRAC $3.6\micron$ closeup image of  
IC~4970, tracing the distribution of old (M0) stars. Contour levels are 
$0.02$ and $0.12$ mJy\,arcsec$^{-2}$. The image field of view is  
$32\farcs4 \times 51\farcs4$ with  
$1\,{\rm pixel} = 0\farcs86 \times 0\farcs86$.
({\it bottom}) {\it Chandra} $0.5-2$\,keV image of IC~4970 with IRAC 
$3.6\micron$ contours from the top panel superposed. The X-ray
emission is concentrated in two X-ray point sources, one at the nucleus and 
one located $2\farcs6$ ($0.68$\,kpc) to the north of the nucleus.
$1\,{\rm pixel} = 0\farcs492 \times 0\farcs492$. }
\label{fig:closeup}
\end{center}
\end{figure}

In the top panel of Figure \ref{fig:closeup} we show a Spitzer IRAC 
$3.6\micron$ image of IC~4970, tracing the stellar distribution in the 
galaxy. For reference, $0.02$ and $0.12$ mJy\,arcsec$^{-2}$ contour 
levels for the $3.6\micron$ emission are also shown. The $3.6\micron$ 
emission seen in the lower  southeast corner of the image is from the 
northern spiral arm of NGC~6872. In the bottom panel of
Fig. \ref{fig:closeup} we show the $0.5-2$\,keV {\it Chandra} X-ray
image of the same field, using the superposed $3.6\micron$ contours 
from the top panel to delineate the stellar extent of IC~4970. 
X-ray emission from IC~4970 is concentrated in two point sources
within  the central $1$\,kpc ($4\farcs$) of the galaxy with one point source 
located at the nucleus  
(${\rm RA}=20^h16^m57^s.56$, ${\rm DEC}=-70^\circ44'59\farcs9$), 
and one $0.68$\,kpc ($2\farcs6$) to the north 
(${\rm RA}=20^h16^m57^s.6$, ${\rm DEC}=-70^\circ44'57\farcs3$). 
The analysis regions we use to study these sources and diffuse
emission in IC~4970 are listed in Table \ref{tab:regions}. 

\begin{figure*}[t]
\begin{center}
\includegraphics[height=3in,width=2.25in,angle=270]{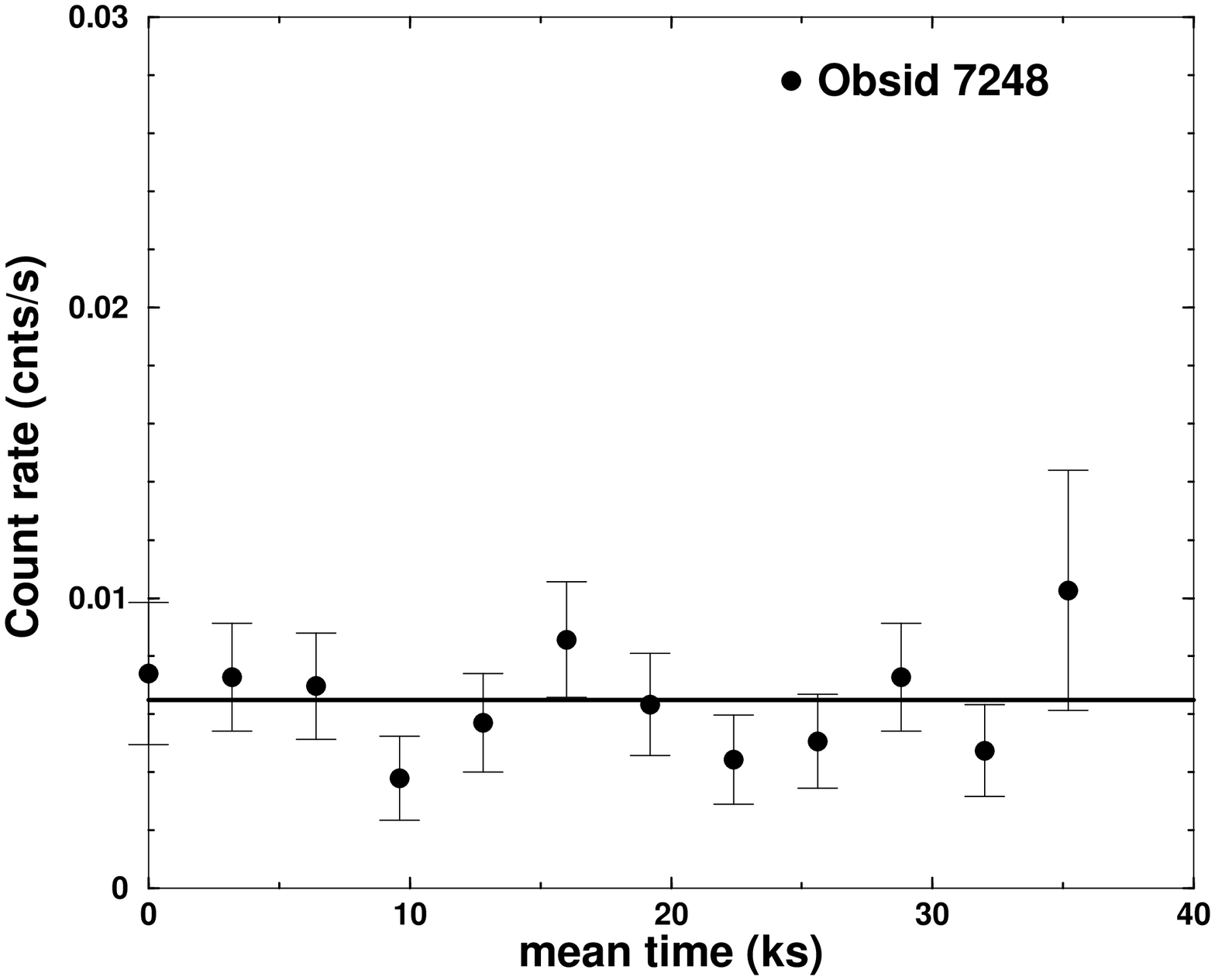}
\includegraphics[height=3.0in,width=2.25in,angle=270]{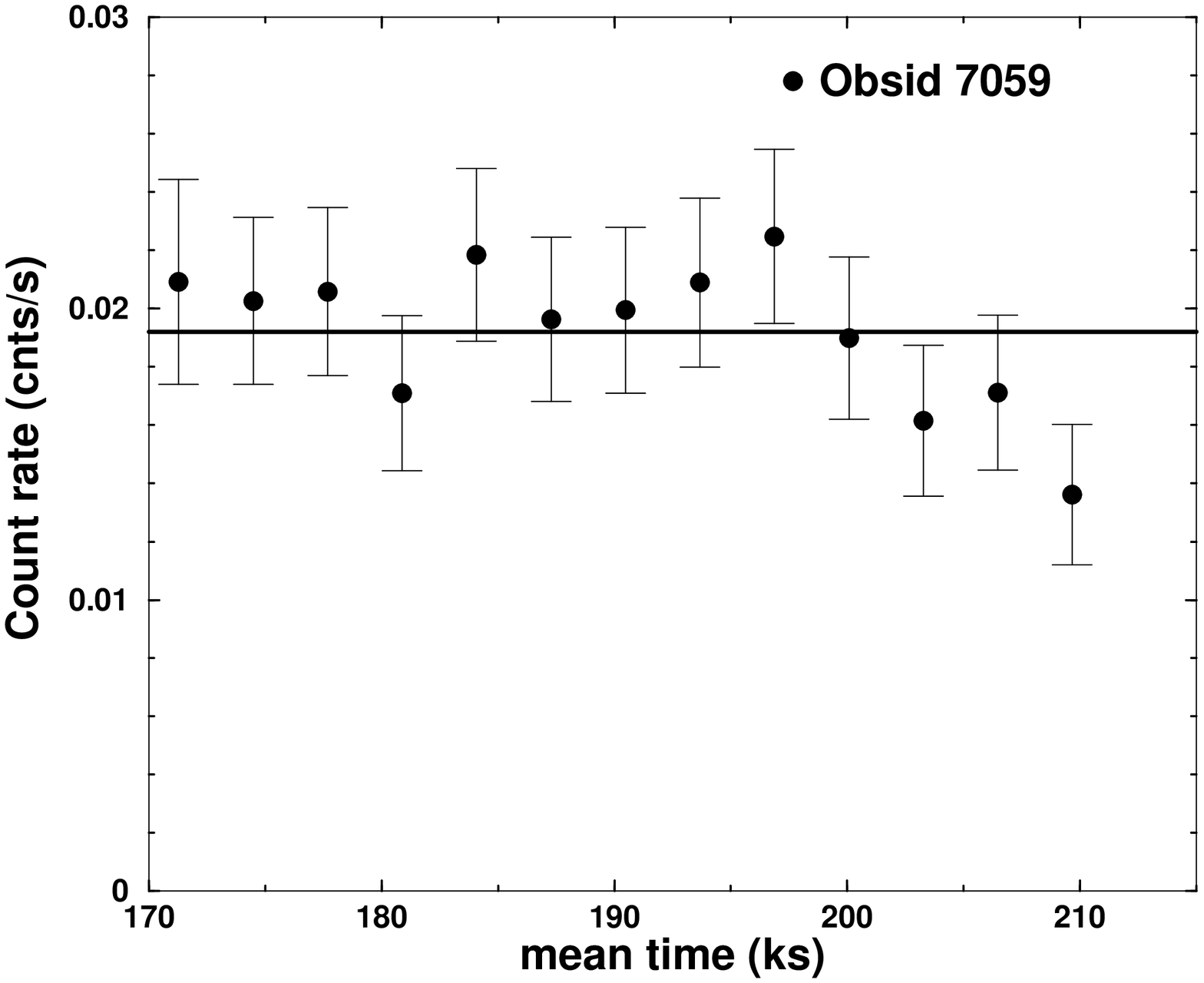}
\caption{{\it Chandra} X-ray light curves for 
an elliptical region with semiaxes of $2\farcs2$ and $1\farcs7$ 
centered on the 
nucleus of IC~4970 in $3.2$\,ks bins, showing a factor of three  
increase in mean count rate over the $\sim 136$\,ks time period
between Obsid 7248 (left panel) and Obsid 7059 (right panel). The mean
time zero in both panels is measured from the start of Obsid 7248.
}
\label{fig:lightcrvs}
\end{center}
\end{figure*}

\subsection{Nuclear X-ray Variability}
\label{sec:vary}

The X-ray luminosity of IC~4970 is dominated by emission from the
nuclear point source. Enhanced nuclear activity is often found as a result of 
gas rich mergers, either in the form of a circumnuclear starburst or
increased AGN activity. To explore the nature of the nuclear source in
IC~4970, we constructed light curves in the $0.5-10$\,keV energy band 
for our two observations, that were taken $136$\,ks apart, using an
elliptical region with semiaxes of $2\farcs2$ and $1\farcs7$ (denoted
`nucleus' in Table \ref{tab:regions}) and 
$3.2$\,ks binning to reduce statistical scatter. Errors were
calculated using the Gehrels method for low count rate data 
(Gehrels 1986). Light
curves for the nucleus, calculated with and without background
subtraction, showed no significant difference. 
Figure \ref{fig:lightcrvs} shows that the nucleus is 
highly variable, 
with the mean count rates, $(6.5 \pm 1.8) \times 10^{-3}$\cnts for
Obsid 7248 (left panel) and $(19.2 \pm 2.5) \times 10^{-3}$\cnts for 
Obsid 7059 (right panel), 
increasing by a factor $\sim 3$  over the $136$\,ks interval
between the observations. To check for any 
systematic variation in count rate between the two observations, we 
used the same energy band and time binning to compute the light curve
for a $30\farcs$ circular region outside IC~4970 (denoted `IGM' in Table
\ref{tab:regions}). We find no statistically significant time
variation between the two observations for region `IGM'. Scaling the 
observed IGM count rate by the ratio of the solid angles for the two 
regions (`nucleus' and `IGM'), 
we find that the mean IGM background count rates for IC~4970's nucleus, 
$(1.86 \pm 0.62) \times 10^{-5}$\cnts and 
$(2.2 \pm 0.75) \times 10^{-5}$\cnts for the two observations, agree 
within their $1\sigma$ uncertainties.

Short time scale variability is not expected 
for circumnuclear starbursts. Thus the observed variability 
strongly suggests that the nucleus of IC~4970 is an AGN. 
  
\subsection{Nuclear X-ray Spectra and Luminosities }
\label{sec:xrayspec}

\begin{figure*}[t]
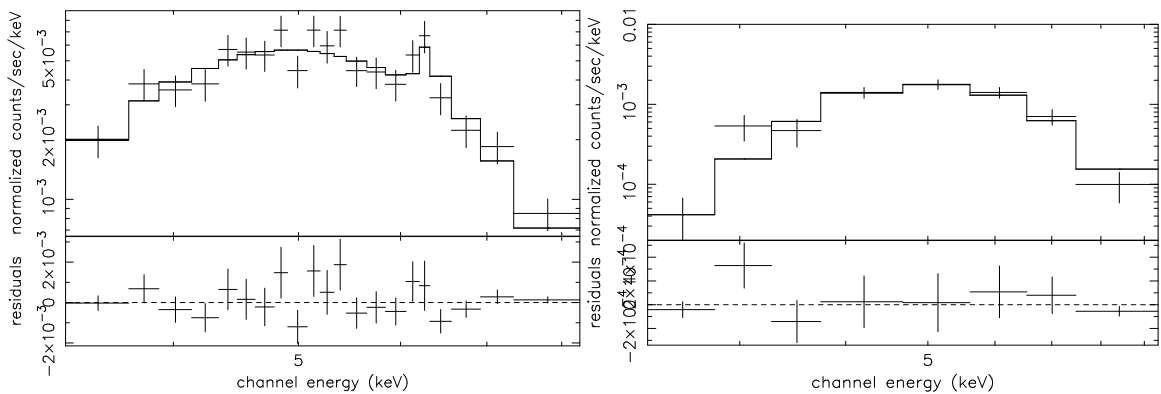

\begin{center}
\includegraphics[height=3in,width=2.0in,angle=270]{f4a.ps}
\includegraphics[height=3in,width=2.0in,angle=270]{f4b.ps}
\caption{ ({\it left}) Spectrum of the nucleus of IC~4970 during Obsid
  7059 (bright state). The spectral model (solid line) is an absorbed
  power law plus Gaussian, with hydrogen column 
  $3 \times 10^{23}$\cms, 
  photon index $1.7$, and a narrow Gaussian  with line energy 
  and redshift of $6.4$\,keV and $0.015728$, respectively. 
({\it right}) Spectrum of the nucleus of IC~4970 during Obsid 7248
  (faint state). The spectral model (solid line) is an absorbed
  power law  with photon index $1.7$ and 
  hydrogen column $3.1 \times 10^{23}$\cms.
  Please note the difference in vertical scales, and 
  each major tick mark in the right residual panel is $2 \times 10^{-4}$.
}
\label{fig:agnspec}
\end{center}
\end{figure*}

We extracted the spectrum of IC~4970's nucleus in the same elliptical
region 
used to construct the light curves, and used a local background
annulus concentric with the nucleus with
inner and outer radii of $4\farcs9$ and $19\farcs7$, respectively,
(denoted `nucleus bkg' in Table \ref{tab:regions}) to
subtract the background.  

\subsubsection{Bright State}
\label{sec:brispec}

We found $683 \pm 26$ net source counts in the 
$0.5-10$\,keV energy band for the bright state (Obsid 7059).
Bright unresolved sources observed with ACIS may be subject to 
pileup, where two
or more photons incident on the same detection cell are recorded as a 
single event with a pulse height amplitude  that is roughly the
sum of the individual event pulse height amplitudes (Davis 2001).
From the observed mean count rate per $3.2$\,s
frame ($0.06$\,cnts\,frame$^{-1}$) for the nucleus in the bright
state, we used PIMMS to estimate a  pileup fraction of $\sim 3\%$ in
our data, such that pileup should not significantly affect our results.
We grouped the spectrum to ensure a minimum of $30$ counts per bin. 
As shown in the left panel of Figure \ref{fig:agnspec}, the spectrum
is highly absorbed and shows a clear $6.4$\,keV Fe K$\alpha$ line. 

We model the spectrum over
the $3-10$\,keV range using a power law model  with an added narrow 
Gaussian to model the Fe K$\alpha$ line. The model is  corrected for 
absorption using 
photoelectric cross sections from Morrison \& McCammon (1983). 
For the narrow Gaussian line profile, 
we fix 
the central line energy at $6.4$\,keV, and redshift at $z = 0.015728$. 
If we allow all other parameters to vary freely, we find a best fit 
hydrogen absorbing column of $2.9^{+0.7}_{-1.0} \times
10^{23}$\cms. Although the central value for the photon index from this
fit ($\Gamma = 1.6$) is consistent with that expected for an AGN,
$\Gamma$  is poorly constrained. 
We thus also fix the photon index $\Gamma$ in our models to lie in the 
range  $1.5-2.0$ typical for highly obscured AGN
(Cappi \etal 2006). The results of our spectral
fits are summarized in Table \ref{tab:nucspec}. 
For $\Gamma = 1.5-2.0$, we find column densities of 
$2.8 - 3 \times 10^{23}$\cms with $10\%$ uncertainties
and Fe K$\alpha$ line equivalent widths of $144-195$\,eV. 
The intrinsic X-ray luminosity of the nucleus in the 
bright state is high, with a $0.5-10$\,keV luminosity of 
$1.7^{+0.6}_{-0.3} \times 10^{42}$\ergs, where the errors reflect 
the uncertainty in the photon index in the spectral models. 

\subsubsection{Faint State}
\label{sec:faintspec}

The spectrum of the faint state (Obsid 7248) is shown in the right  
panel of Figure \ref{fig:agnspec}.  
There are $196 \pm 14$ net source counts for the faint state 
in the $0.5-10$\,keV energy band with only 
$15.6 \pm 3.9$ net source counts below $2$\,keV. 
We grouped the spectrum using a predefined 
grouping for very faint sources that results in channels of approximately
constant logarithmic width, and 
restricted our spectral fits to the $2-10$\,keV energy band. We again model 
the spectrum using absorbed power law models with fixed photon index 
$\Gamma$ in the range $1.5-2.0$.  The
$\chi^2/{\rm dof}$ in the model fits to the faint state favor 
steeper $\Gamma \sim 1.7-2.0$, as expected for a Seyfert $2$ nucleus. 
The hydrogen absorption column for 
these models is $(2.9 -3.4) \times 10^{23}$\cms, with $\sim \pm 20\%$ 
uncertainties from the spectral fits, in agreement with the spectral 
model fits for the bright state (See Table \ref{tab:nucspec}).
The $0.5-10$\,keV X-ray 
luminosity of the nucleus in the 
faint state is $0.6^{+0.2}_{-0.1} \times 10^{42}$\ergs, 
where the uncertainties
are due to the uncertainty in the spectral model $\Gamma$. 

Thus, with these data, we see no significant difference 
between the bright and faint states in either the 
power law slope or hydrogen absorption column. We find only that the
X-ray luminosity is reduced by a factor $2.9$. 
 
\subsubsection{Soft Emission in the Nucleus}
\label{sec:thermal}

There are a total $41 \pm 6.4$ source counts from the nuclear region 
in the $0.5-2$\,keV energy 
band for the combined observations. 
If we assume that this soft emission is solely due to 
thermal emission from hot gas in the nuclear region outside the 
obscuring torus, we can place an upper limit on the amount of hot gas 
within $\sim 0.5$\,kpc of the nucleus. 
Using Anders \& Grevasse (1989) abundances and assuming solar metallicity,  
the spectrum of this soft emission
is consistent with an absorbed thermal APEC model (Smith \etal 2001) 
with absorbing column 
$N_{\rm H} =(8 \pm 4) \times 10^{21}$\cms and temperature 
$kT =0.6^{+0.8}_{-0.4}$\,keV ($\chi^2/{\rm dof} = 6.1/7$), and an estimated 
$0.5-10$\,keV intrinsic luminosity of $9 \times 10^{39}$\ergs. 
From the spectral model normalization and assuming uniform filling, we
estimate an electron density and hot gas mass in the nuclear region of 
$\lesssim 0.2$\cmc and $\lesssim 3 \times 10^6\Ms$, respectively.

\subsection{IRAC Mid-Infrared Colors of the Nucleus}
\label{sec:iraccol}

\begin{figure}[t]
\begin{center}
\includegraphics[height=2.2in,width=3in]{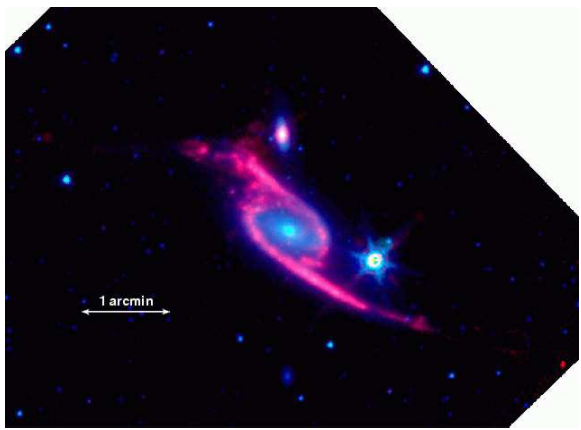}
\caption{{\it Spitzer} mid-infrared composite image of NGC~6872 and IC~4970
  taken in the IRAC wavebands. Blue (green) shows the $3.6
  (4.5)\micron$ emission from old stars, respectively, and red shows the 
 $8\micron$ nonstellar emission (see Machacek \etal 2007a, in prep.).
 North is up and east is to the left.}
\label{fig:n6872irac}
\end{center}
\end{figure}

Heavily absorbed nuclear regions of galaxies are typically bright
in the mid-infrared. Figure \ref{fig:n6872irac} shows a composite 
IRAC mosaic of the NGC~6872/IC~4970 system. Blue and green denote 
$3.6\micron$ and $4.5\micron$ emission, respectively, that trace old stars 
in the galaxies. The $8\micron$ emission is a combination of starlight
and `nonstellar' emission sources such as warm dust and AGN. We model 
the stellar contribution by averaging background-subtracted 
$3.6\micron$ and $4.5\micron$ emission maps that have each
been aperture corrected and scaled to the $8\micron$ waveband
using the mid-infrared colors of M0III stars. We then subtract 
this model from the 
background-subtracted $8\micron$ mosaic to produce a map of
`nonstellar' $8\micron$ emission, shown in red (Pahre \etal 2004; 
Machacek \etal 2007a, in preparation). Figure \ref{fig:n6872irac} shows 
that the nonstellar $8\micron$ emission in IC~4970 is concentrated in
the inner $4\farcs-5\farcs$ ($\sim 1$\,kpc) of the galaxy, with 
the peak of the nonstellar emission at the nucleus of IC~4970, 
qualitatively consistent with a heavily obscured active nucleus 
(Pahre \etal 2004). To check whether the $8\micron$ nonstellar
emission is extended or consistent with an unresolved point source, 
we extracted $2\farcs2$ wide vertical and horizontal cuts through the
$8\micron$ nonstellar surface brightness distribution at the nucleus. 
The full width at half-maximum of the resulting projected surface 
brightness distribution in each case was $\sim 3\farcs$, a factor
$1.5$ larger than that of the IRAC point spread function in this
waveband, suggesting part of the emission arises from a roughly spherical
distribution of warm dust in the central region of the galaxy outside 
the luminous accretion disk of the AGN. The bright patches of nonstellar
$8\micron$ emission from polycyclic aromatic hydrocarbons (warm dust) 
along NGC~6872's northern arm and at the `knee' and `bridge', that
trace the collision path of IC~4970 through NGC~6872, also show that 
IC~4970 has recently tidally interacted with a region in the spiral
that is rich in cold gas and dust, such that the spiral galaxy
NGC~6872 could be one source for the obscuring material now found in 
the less massive, tidal perturber IC~4970.

We perform fixed aperture photometry on the background-subtracted
mosaiced-images in all four IRAC wavebands ($3.6$, $4.5$, $5.8$, and 
$8.0\micron$), using a $2\farcs44$ circular aperture and  background
annulus with inner, outer radii of $2\farcs44$, $7\farcs32$, respectively, 
centered on the nucleus of IC~4970. The data were aperture corrected 
using point source aperture corrections for each IRAC waveband 
appropriate for the chosen aperture and background geometries, and the 
resulting flux densities converted to Vega relative 
magnitudes. For completeness, the aperture correction factors and 
Vega relative zeropoints used for each IRAC channel are listed 
in Table \ref{tab:photfac}. Our results for the IRAC-band flux densities  
(magnitudes) of IC~4970's nucleus are given in Table \ref{tab:nucmir}. 

We find IRAC mid-infrared colors of $[3.6]-[4.5] = 0.17$  and 
$[5.8]-[8.0] = 1.59$. These mid-infrared colors fall 
outside the mid-infrared color selection criteria used by 
Stern \etal (2005) to 
separate active galaxies from normal galaxies. The $[3.6]-[4.5]$ color
of IC~4970's nucleus is too blue, falling below the color-color
selection `wedge' shown in their Figure 1, but the $[5.8]-[8.0]$ color 
of IC~4970's  nucleus is far redder than the zero redshift endpoint 
of the evolutionary track for the  normal S0/Sa galaxy NGC~4429, also 
shown in that figure. 
However, the Stern \etal selection `wedge' was  based on a photometrically
identified sample of bright quasars and Seyfert $1$ galaxies, rather 
than highly obscured Seyfert $2$s.  Given that $60\%$ of the 
narrow line AGN identified in their sample also fall outside the 
color-color 
selection 'wedge', these criteria likely do not apply to IC~4970. 
Lacy \etal (2004) used quasars from the Sloan Digital Sky Survey
to empirically define a selection region in the mid-infrared color-color
space defined by the $8.0\micron/4.5\micron$ and  $5.8\micron/3.6\micron$  
flux density ratios. Objects red in both colors are likely AGN. 
These mid-infrared colors for the nucleus of IC~4970 (
${\rm log}(S_{5.8}/S_{3.6}) =  0.13$ and ${\rm log}(S_{8.0}/S_{4.5}) =
0.6$) are both red and are consistent with the
colors of obscured AGN candidates in the Lacy \etal sample 
(See Fig. 1 in Lacy \etal 2004).  

Thus both the X-ray properties (spectrum and variability) and
mid-infrared colors suggest IC~4970 hosts a highly obscured 
Seyfert $2$ nucleus.

\subsection{Near-Nuclear X-ray  Source}
\label{sec:ULX}

We use a $1\farcs3$ circular source region and the same annular 
background region that was used for the nucleus (`north source' and 
`nucleus bkg', respectively, in Table \ref{tab:regions}) 
to measure the net X-ray source counts 
in the combined $70.2$\,ks observation from  the near-nuclear X-ray  
source, located  $0.68$\,kpc to the north of the AGN (see the right 
panel of Figure \ref{fig:closeup}). 
Since the near-nuclear X-ray source
is separated from the nucleus by only $\sim 2\farcs6$ and the nucleus is
located  $3\farcm19$ from the ACIS-I aim point for this observation, 
the AGN and near-nuclear point source are resolved in the $0.5-2$\,keV
energy band, but are not resolved in the $2-10$\,keV energy band. 
We find a total of $37.5 \pm 6.2$ net source
counts in the $0.5-2$\,keV energy band for the northern, near-nuclear 
 X-ray source and, to avoid contamination from the nucleus, use only the  
$0.5-2$\,keV count rate to determine its luminosity. 
Assuming Galactic absorption of $5 \times
10^{20}$\cms, and absorbed power law spectral models with photon
indices $\Gamma = 1.7 \pm 0.3$, typical for luminous X-ray point sources
(see, e.g. Colbert \etal 2004; Swartz \etal 2004), we find a 
$0.5-10$\,keV luminosity for 
the near-nuclear X-ray point source of $3.5^{+1.8}_{-1.2} \times
10^{39}$\ergs, where
the errors reflect both the range of photon indices considered in 
our spectral models and the $1\sigma$ statistical uncertainty in the 
$0.5-2$\,keV count rate. Thus the near-nuclear X-ray point 
source is an ultra-luminous X-ray source. Our estimate for the 
luminosity of the near-nuclear  X-ray source should be considered a 
lower bound on its intrinsic luminosity, since $8\micron$ nonstellar
emission in the center of IC~4970 suggests that obscuring material 
may extend outside the nucleus (see \S\ref{sec:iraccol}), and 
ultra-luminous X-ray sources, on average, tend to be moderately absorbed 
($N_{\rm H} \sim 2 \times 10^{21}$\cms; Swartz \etal 2004).

\subsection{Diffuse Emission Outside the Central Region}
\label{sec:diffuse}

To estimate the amount of hot gas in IC~4970 outside the nucleus, 
we measure the X-ray emission from  an elliptical region, based on  
the distribution of starlight in the image of IC~4970 shown in 
the right  panel of Figure
\ref{fig:pavogroup}, excluding a central ellipse containing the two
point sources near the nucleus 
(denoted `full extent' and  `central' in Table \ref{tab:regions}). The
local background is again the nearby $30\farcs$ circular region `IGM'
in Table \ref{tab:regions}. 
X-ray emission outside the central region is weak with 
only $28.3 \pm 6.4$ net source counts in the $0.5-2$\,keV energy band, 
for the combined $70.2$\,ks observation. We estimate the
amount of this emission that is the result of low mass X-ray
binaries in IC~4970 using the low mass X-ray binary cumulative 
luminosity function for early type galaxies from Gilfanov (2004),
normalized by the stellar mass determined from K-band 
photometry of the  $2$ Micron All Sky Survey image of the same 
region. We model the excess 
$0.5-2$\,keV emission with a Raymond-Smith thermal  plasma model, 
assuming a gas temperature of $0.54$\,keV and metallicities 
of $0.4-1\Zs$, typical for the interstellar medium of early type 
galaxies (see, e.g. Machacek \etal 2004, 2006), and 
Galactic absorption of 
$5 \times 10^{20}$\cms (Dickey \& Lockman 1990). 
We find a total $0.5-2$\,keV
luminosity outside the central region of IC~4970 of $10^{39}$\ergs,
with  $6.8 \times 10^{38}$\ergs ($66\%$) from the thermal component and 
$3.5 \times 10^{38}$\ergs ($34\%$) from low mass X-ray binaries. Using the
Raymond-Smith spectral model normalization to estimate the mean
electron density and gas mass outside IC~4970's central region, 
we find (for uniform filling) mean electron densities  of 
$(2-3) \times 10^{-3}$\cmc and hot gas masses of  
$\sim 1-2 \times 10^7\Ms$, where the ranges  
reflect uncertainties in the assumed metal abundance and geometry. 
This is a factor $\gtrsim 50$ less than the amount of cold HI gas that may 
be associated with the galaxy (Horellou \& Koribalski 2007).
Thus either the gravitational potential of IC~4970 is too shallow to hold a
significant hot gas halo or, alternatively, the hot gas 
has been stripped during the collision of IC~4970 with the spiral
galaxy NGC~6872 and/or by its interaction with the Pavo group 
intra-group medium 
during the supersonic passage of the NGC~6872/IC~4970 galaxy pair 
through the Pavo group core. 

\section{Fueling the Central Black Hole}
\label{sec:discuss}

\subsection{Determining the Black Hole Mass}

The mass of the central black hole in galaxies is tightly correlated
with the absolute K-band luminosity of the galaxy bulge,  
\begin{equation} 
 {\rm log}M_{\rm BH} = 8.21 + 1.13{\rm log}(L_K - 10.9)
\label{eq:bhmass}
\end{equation}
 where $M_{\rm BH}$ is the mass of the black hole in solar masses and 
 and $L_K$ is the absolute K-band luminosity of the galaxy's bulge 
in solar luminosities (Marconi \& Hunt 2003). For early type galaxies, 
like IC~4970,  the bulge luminosity in
 eq. \ref{eq:bhmass} is replaced by the total K-band luminosity of the 
galaxy as a whole. Thus we use photometry of the `full
 extent' source region from Table \ref{tab:regions}, with the 
  `IGM' region as a local background,  on $2.2\micron$ images 
 from the $2$ Micron All Sky Survey to 
 measure the K-band luminosity of IC~4970. Since it is likely that  
 IC~4970  and NGC~6872 lie near the core of the Pavo group  
(Machacek \etal 2005), such that the luminosity distance is
 $55.5$\,Mpc, we find ${\rm log}(L_K/L_\odot) = 10.46$ and the central
black hole mass of $5 \times 10^7\Ms$. If instead we use the
 B-band mass-to-light ratio ($ M_{\rm dyn}/L_B = 10$ ) for IC~4970,
 that was needed in simulations to reproduce the tidal features
 observed in the spiral galaxy NGC~6872 (Mihos 1993; 
Horellou \& Koribalski 2007), we find a dynamical mass of 
$\sim 10^{11}\Ms$ within a B-band effective radius of $11\farcs$
 ($3$\,kpc), implying a line-of-sight velocity dispersion of $131$\kms. 
Then from the $M_{\rm BH} - \sigma$ relationship (Gebhardt \etal 2000; 
Ferrarese \& Merritt 2000), we infer a central black hole mass for 
IC~4970 of $2-3 \times 10^7\Ms$, a factor $\sim 2$ smaller than that 
obtained from the $L_K-M_{\rm BH}$ relationship, but consistent within its 
$\pm 0.5$\,dex scatter (Hardcastle \etal 2007). 

\subsection{Mass Accretion Rates and Modes}

From the X-ray properties of the nuclear region of 
IC~4970, we can estimate the accretion power available from hot gas 
in the nuclear region, and compare the available accretion power from 
hot gas to the luminosity of the nucleus to constrain the 
accretion mode of the AGN. For a $5\times 10^7\Ms$
black hole the Eddington luminosity is 
$L_{\rm Edd} = 6.3 \times 10^{45}$\ergs, such that for the 
bright state, 
$L_{\rm X}/L_{\rm Edd} = 2.7^{+1.0}_{-0.5} \times 10^{-4}$,
consistent with  $L_{\rm X}/L_{\rm Edd}$ observed for heavily
absorbed, accretion powered radio galaxies (Evans \etal 2006).
Thus, even in the bright state, the radiative power is 
sub-Eddington.   The mass accretion rate needed to produce the observed 
X-ray luminosity is given by 
 $\dot{m} =  L_{\rm X}/(\eta c^2)$, where $\eta$ is the
  radiative efficiency and  $c$ is the speed of light. 
Assuming a $10\%$ radiative efficiency, the mass 
accretion rate needed to power the AGN in IC~4970 is 
$\dot{m} = 1.9^{+0.6}_{-0.3} \times 10^{22}$\,g\,s$^{-1}$. This should be 
considered a lower limit on the required mass accretion rate, since 
for accretion powered AGN, the bolometric luminosity is likely a 
factor $\sim 3-10$ higher (Elvis \etal 1994) than $L_{\rm X}$, and radiative
efficiencies $\lesssim 0.01$ may be more common (Evans \etal 2006).

The accretion rate available from the Bondi accretion of hot gas onto
the central black hole is given by  
\begin{equation}
 \dot{M}_{\rm Bondi} = \frac{\pi G^2 M^2_{\rm BH}\rho}{c_s^3}
\label{eq:bondi}
\end{equation}
where $G$ is the gravitational constant, $M_{\rm BH}$ is the black
hole mass, $c_s$ is the sound speed and $\rho$ is the gas density 
at the Bondi radius $r_A = 2GM_{\rm BH}/c_s^2$ 
(see, e.g., Allen \etal 2006). In our case the Bondi radius is small 
($\sim 3$\,pc), so we use the limits on the hot gas near the nucleus 
derived from assuming all of the excess $0.5-2$\,keV emission in the 
nuclear region was thermal (see \S\ref{sec:xrayspec}). For hot gas 
 with temperature $kT \sim  0.6$\,keV and electron density 
$n_e \sim 0.2$\cmc accreting onto a 
$5\times 10^7\Ms$ black hole,  we find a Bondi accretion rate of 
$8.4 \times 10^{20}$\,g\,s$^{-1}$. This is less than $5\%$ of the 
lower limit on the 
mass accretion rate $\dot{m}$ needed to power the AGN. 
 
The available Bondi accretion power is very sensitive to the 
temperature of the hot gas  through the dependence of eq.
\ref{eq:bondi} on the gas sound speed, with lower gas temperatures 
allowing higher accretion rates. However, even for the $90\%$ 
confidence lower limit on the gas temperature ($kT = 0.2$\,keV) 
for hot gas in the nuclear region, the available mass accretion rate 
($6.8 \times 10^{21}$\,g\,s$^{-1}$) is still a factor of three too small 
to power the observed X-ray luminosity and likely an 
order of magnitude too small to provide the total bolometric power for 
the AGN. Furthermore, 
if we have overestimated the black hole mass by a factor two, as suggested 
by previous numerical simulations of the interaction of IC~4970 with 
the larger spiral galaxy NGC~6872 (Mihos \etal 1993, 
Horellou \& Koribalski 2007), the 
Bondi accretion rates quoted above would be reduced by an additional 
factor of four. 

Thus the most likely explanation for the nuclear activity in 
IC~4970 is that the active nucleus was triggered 
by the accretion of cold (HI) gas and dust driven into its center as a
result of IC~4970's  ongoing interaction with the spiral galaxy NGC~6872.  
Since numerical simulations of the NGC~6872/IC~4970 system 
by Horellou \& Koribalski (2007) suggest that IC~4970 will accrete
significant amounts of matter during this interaction, some, if not
all, of the dust and cold gas fueling IC~4970's nucleus may have been 
acquired by IC~4970 during its passage through the northern spiral arm 
of its much larger, gas-rich spiral partner.
Such interactions would have been frequent between
galaxies in groups at high redshift, when the fraction of gas-rich 
spirals was high, thereby promoting AGN activity and black hole growth 
in lower mass galaxies, that are unable to retain significant hot gas halos
because of their shallow gravitational potentials.

\section{Conclusions}
\label{sec:conclude}

In this paper we use {\it Chandra} X-ray and {\it Spitzer} IRAC 
mid-infrared observations to probe nuclear activity in
IC~4970, the interacting companion galaxy to the large spiral galaxy
NGC~6872 in the Pavo galaxy group. We find the following: 
\begin{itemize}
\item{X-ray emission from the nucleus of IC~4970 is time variable with
  the mean $0.5-10$\,keV count rate increasing by a 
  factor of three on $\sim 100$\,ks time scales. This strongly suggests 
  IC~4970's nucleus hosts an AGN rather than a compact nuclear starburst.}
\item{The X-ray spectrum of IC~4970's nucleus in the bright state
  shows a clear Fe K$\alpha$ line and is well described by an 
  absorbed power law plus narrow 
  Gaussian line model with absorption column 
  $N_{\rm H} = 3 \times 10^{23}$\cms,
  fixed photon indices $\Gamma = 1.5 - 2$, and an Fe K$\alpha$ line equivalent 
  width of $144-195$\,eV. The $0.5-10$\,keV X-ray 
  luminosity of the nucleus in the bright state is
  $1.7^{+0.6}_{-0.3} \times 10^{42}$\ergs 
  and in the
  faint state is  $0.6^{+0.2}_{-0.1} \times 10^{42}$\ergs.  
  With these data, we find no significant difference in the power law
  slope or intrinsic absorption between the bright and faint nuclear
  states.}
\item{Nonstellar $8\micron$ emission is concentrated in the central 
  $\sim 1$\,kpc ($4\farcs-5\farcs$) of IC~4970, consistent with 
  high obscuration in this region.  The mid-infrared colors of the nucleus, 
  ${\rm log}(S_{5.8}/S_{3.6})=0.13$, ${\rm log}(S_{8.0}/S_{4.5})=0.6$, 
  are consistent with the mid-infrared colors expected for highly obscured 
  AGN candidate sources. Thus the X-ray and mid-infrared properties of the 
  nucleus of IC~4970 suggest that it is a highly obscured Seyfert $2$. }
\item{Assuming that the $0.5-2$\,keV X-ray emission observed in the nuclear
  region is all thermal, the spectrum is
  consistent with an absorbed $0.6^{+0.8}_{-0.4}$\,keV APEC model with  
  solar metallicity, a hydrogen absorbing column of 
  $8 \pm 4 \times 10^{21}$\cms, and intrinsic $0.5-10$ luminosity of 
   $9 \times 10^{39}$\ergs. For uniform filling, the inferred mean 
  electron density and hot gas mass in the nuclear region 
  are $\lesssim 0.2$\cmc and 
  $\lesssim 3 \times 10^{6}\Ms$, respectively.}
\item{An ultra-luminous X-ray source with $0.5-10$\,keV intrinsic 
  luminosity of $\gtrsim 3.5^{+1.8}_{-1.2} \times 10^{39}$\ergs
  is found $0.68$\,kpc from IC~4970's nucleus.} 
\item{ Little diffuse X-ray emission is observed outside
  the nuclear region. Assuming $0.54$\,keV interstellar galaxy gas, 
  the X-ray luminosity 
  of diffuse gas outside the central $1$\,kpc of the galaxy is 
  $\sim 7 \times 10^{38}$\ergs and gas mass is $\sim 1-2 \times
  10^7\Ms$. Either the gravitational potential of IC~4970 is too
  shallow to retain a significant hot gas halo, or that halo has been
  stripped by the interaction of IC~4970 with NGC~6872 and the
  Pavo group intra-group medium.}
\item{From the correlation between black hole mass and K-band
  luminosity, we expect IC~4970 to host a $5 \times 10^7\Ms$ black hole. 
  Bondi accretion of hot gas in the nuclear region of IC~4970 onto the
  black hole can account for only $\sim 5\%$ of the observed X-ray 
  luminosity of the AGN, such that the dominant power source for the
  active nucleus is likely the accretion of dust and cold gas driven into 
  IC~4970's nuclear region by the galaxy's ongoing interaction with
  the gas-rich spiral galaxy NGC~6872.}
   
\end{itemize}
 

\acknowledgements

Support for this work was provided, in part, by the National Aeronautics
and Space Administration (NASA) through {\it Chandra} Award Number
GO6-7068X issued by the Chandra X-ray Observatory Center, which is
operated by the Smithsonian Astrophysical Observatory for and on
behalf of NASA under contract NAS8-03060, by NASA through an award
issued by JPL/Caltech, and by the Smithsonian Institution. This work
is also based in part on observations made with 
the Spitzer Space Telescope, which is operated by the Jet Propulsion 
Laboratory, California Institute of Technology under a contract with NASA. 
This work is based in part on the IRAC `post-Basic Calibrated Data' 
processing software IRACproc developed by Mike Schuster, Massimo
Marengo and Brian Patten at the Smithsonian Astrophysical Observatory.
This work has made use of data products from the Two 
Micron All Sky Survey, which is a joint project of the University of 
Massachusetts and the Infrared Data Analysis and Processing 
Center/California Institute of Technology, funded by NASA 
and the National Science Foundation, and has also used the NASA/IPAC 
Extragalactic Database (NED), which is operated by JPL/Caltech, 
under contract with NASA.   
 We wish to thank Massimo Marengo for help with IRACproc.


\begin{small}

\end{small}

\begin{deluxetable}{lcccc}
\tablewidth{0pc}
\tablecaption{IC~4970 Analysis Regions\label{tab:regions}}
\tablehead{
\colhead{Source}& \colhead{Geometry}&\colhead{Center}&\colhead{Radii} & \colhead{PA} \\
& & RA, DEC  & arcsec,arcsec   & deg  }
\startdata
Nucleus & ellipse &$20\,\,16\,\,57.6$, $-70\,\,44\,\,59.9$ & $2.2$,$1.7$& $0$
\\
Nucleus Bkg & annulus &$20\,\,16\,\,57.6$, $-70\,\,44\,\,59.9$ &
$19.7$,$4.9$& \ldots \\
North Source & circle & $20\,\,16\,\,57.6$, $-70\,\,44\,\,57.2$ & $1.3$ &
\ldots\\
Central & ellipse & $20\,\,16\,\,57.8$, $-70\,\,44\,\,59.5$ & $4.4$,$3.9$ &$0$\\
Full Extent & ellipse & $20\,\,16\,\,57.2$, $-70\,\,44\,\,59.0$ & $18$,$7.5$ & $96$\\
IGM & circle & $20\,\,16\,\,48.2$, $-70\,\,44\,\,32.1$ & $30$ & \ldots \\
\enddata
\tablecomments{ Columns are (1) region label, (2) region geometry, 
(3) region center in J2000.0 coordinates, (4) radius (circle),
  outer, inner radii (annulus), and major, minor semiaxes (ellipse), and (5)
  position angle measured counterclockwise from west for 
  elliptical regions.
 }
\end{deluxetable}

\begin{deluxetable}{lcccc}
\tablewidth{0pc}
\tablecaption{IC~4970 Nuclear X-ray Spectral Fits\label{tab:nucspec}}
\tablehead{\colhead{State} & \colhead{$\Gamma$} &\colhead{$N_{\rm H}$} &
  \colhead{Fe EqWidth} &\colhead{$\chi^2/{\rm dof}$} \\
 & & $10^{23}$\cms & eV & }
\startdata
Bright &  & & & \\
       & $1.5$ & $2.8 \pm 0.3$       & $144$  & $17.14/18$ \\
       & $1.7$ & $3.0^{+0.3}_{-0.4}$ & $171$  & $17.5/18$  \\
       & $1.8$ & $3.1^{+0.3}_{-0.4}$  & $179$  & $17.8/18$  \\
       & $2.0$ & $3.2 \pm 0.3$       & $195$  & $18.6/18$ \\
Faint  &  &  &  &  \\
       & $1.5$ & $2.9^{+0.6}_{-0.5}$ &\ldots  & $6.87/6$ \\
       & $1.7$ & $3.1^{+0.6}_{-0.5}$ &\ldots  & $6.36/6$  \\
       & $1.8$ & $3.2^{+0.6}_{-0.5}$ &\ldots  & $6.16/6$ \\ 
       & $2.0$ & $3.4 \pm 0.6$       &\ldots  & $5.86/6$  \\
\enddata
\tablecomments{Spectral models are an absorbed power law for the faint state
  (Obsid 7248) and an absorbed power law plus narrow Gaussian, with line
  energy ($6.4$\,keV)
  and redshift ($0.015728$) fixed, for the bright state. 
  The photon index $\Gamma$ was also 
  not allowed to vary in the fits. Spectral models were fit over the 
  $3-10$\,keV ($2-10$\,keV) for the bright (faint) state,
  respectively. Errors are $90\%$ confidence levels.
}
\end{deluxetable}

\begin{deluxetable}{ccc}
\tablewidth{0pc}
\tablecaption{Mid-Infrared Photometry Parameters\label{tab:photfac}}
\tablehead{\colhead{waveband} & \colhead{aperture$^a$} &
   \colhead{zeropoint$^b$}  \\
  $\micron$ & correction & Jy }
\startdata
 $3.6$  & $1.213$ & $280.9$ \\
 $4.5$  & $1.234$ & $179.7$ \\
 $5.8$  & $1.379$ & $115.0$ \\
 $8.0$  & $1.584$ & $64.1$  \\
\enddata
\tablecomments{ $^a$ Infrared Array Camera Data Handbook 3.0 (2006), 
 http://ssc.spitzer.caltech.edu/irac/dh/iracdatahandbook3.0.pdf, p. 53. 
 $^b$ Vega relative zeropoints from Reach \etal (2005).
}
\end{deluxetable}

\begin{deluxetable}{ccc}
\tablewidth{0pc}
\tablecaption{IC~4970 Nucleus Mid-Infrared Photometry\label{tab:nucmir}}
\tablehead{\colhead{waveband} & \colhead{flux density} &
   \colhead{magnitude}  \\
  $\micron$ & mJy &  }
\startdata
 $3.6$  & $4.85$  & $11.91$  \\
 $4.5$  & $3.64$   & $11.74$  \\
 $5.8$  & $6.53$  & $10.62$  \\
 $8.0$  & $15.78$  & $9.02$  \\
\enddata
\tablecomments{Results are for a fixed $2\farcs44$  radius circular
  aperture centered on the nucleus of IC~4970 
  ($20^h16^m57^s.4$, $-70^\circ45'\,00\farcs1$) with a 
  concentric background 
  annulus whose (inner, outer) radii are  ($2\farcs44$, $7\farcs32$). 
  Photometric corrections and zeropoints are listed in Table 
  \protect\ref{tab:photfac}. IRAC photometric uncertainties 
  are $\sim 5\%$. 
  }
\end{deluxetable}

\vfill
\eject
\end{document}